\documentclass[aps,twocolumn,superscriptaddress,nofootinbib,floatfix]{revtex4-2}

\usepackage{graphicx}
\usepackage{bm}
\usepackage{amsmath}
\usepackage{amssymb}
\usepackage{physics}
\usepackage[colorlinks=true,linkcolor=blue,citecolor=blue,urlcolor=blue]{hyperref}
\usepackage{changes}
\usepackage{soul}
\soulregister{\cite}{7}
\soulregister{\ref}{7}  
\usepackage{xcolor}

\graphicspath{{figs/}}
\DeclareMathOperator{\sinc}{sinc}
\newcommand{\TE}[1]{\mathrm{TE}_{#1}}

\begin{document}

\title{A Near-Cutoff Waveguide Haloscope for sub-meV Dark Matter}

\author{Chuan-Yang Xing}
\thanks{Corresponding authors: \href{mailto:cyxing@upc.edu.cn}{cyxing@upc.edu.cn}}
\affiliation{College of Science, China University of Petroleum (East China), Qingdao 266580, China}

\author{Bin Zhu}
\thanks{Corresponding authors: \href{mailto:zhubin@mail.nankai.edu.cn}{zhubin@mail.nankai.edu.cn}}
\affiliation{School of Physics, Yantai University, Yantai 264005, China}

\begin{abstract}
We propose a near-cutoff parallel-plate waveguide haloscope for sub-meV dark matter.
The concept retains the large-area openness of a dish antenna while providing cavity-like field enhancement through slow-wave response and coherent accumulation, without relying on a closed standing-wave resonance.
For a copper waveguide, the projected dark photon sensitivity reaches $\varepsilon\simeq2.1\times10^{-15}$ near $m_{A'}\simeq 0.1\,\mathrm{meV}$.
With an external magnetic field, the same transducer can approach QCD axion parameter space.
The waveguide haloscope highlights a sensitive and scalable route toward future sub-meV bosonic dark matter searches.
\end{abstract}

\maketitle

\section{Introduction}

Dark matter (DM) provides one of the strongest empirical indications of physics beyond the Standard Model, yet its microscopic identity remains unknown. Although much of the traditional search program has focused on weakly interacting massive particles (WIMPs)~\cite{Goodman:1984dc}, compelling alternatives include ultralight bosonic DM candidates such as axions~\cite{Preskill:1982cy,Abbott:1982af,Dine:1982ah} and dark photons~\cite{Holdom:1985ag,Nelson:2011sf,Arias:2012az}.
Both candidates arise naturally in broad classes of extensions of the Standard Model: axions are tied to symmetry structure~\cite{Peccei:1977hh,Weinberg:1977ma,Wilczek:1977pj} and often appear in string-motivated constructions~\cite{Arvanitaki:2009fg}, while dark photons emerge naturally from hidden-sector $U(1)$ gauge symmetry and communicate with the visible sector through kinetic mixing~\cite{Holdom:1985ag,Nelson:2011sf,Arias:2012az}.

Within this broader program, the high-frequency / sub-meV frontier is a particularly compelling target. For axions, post-inflation Peccei--Quinn scenarios, together with lattice-QCD input and the dynamics of strings and string-wall collapse, motivate a high-frequency QCD axion DM region in the tens-to-hundreds of $\mu{\rm eV}$ range~\cite{Borsanyi:2016ksw,Gorghetto:2020qws,Buschmann:2021sdq}.
A parametric-resonance production scenario likewise points to an upward shift of the QCD axion DM window toward $10^{-4}$--$10^{-3}\,{\rm eV}$~\cite{Pirzada:2026npl}.
For dark photons, inflationary fluctuations of a massive vector, especially the longitudinal mode, provide a concrete sub-meV target when the abundance is tied to a high inflationary scale~\cite{Graham:2015rva}.
The sub-meV range can also be populated through other nonthermal production mechanisms, including tachyonic production from axion oscillations~\cite{Agrawal:2018vin,Co:2018lka}, production at the end of inflation~\cite{Bastero-Gil:2018uel}, and cosmic-string emission~\cite{Long:2019lwl}.
Once produced, such ultralight bosonic DM is naturally described today as a highly occupied, locally phase-coherent classical field~\cite{Sikivie:2020zpn,Fabbrichesi:2020wbt,Chadha-Day:2021szb}.

However, detecting the sub-meV frontier remains technically challenging.
Closed microwave cavities~\cite{Sikivie:1983ip,DePanfilis:1987dk,Hagmann:1990tj,ADMX:2025vom,HAYSTAC:2024jch,Bae:2024kmy,Quiskamp:2024oet,QUAX:2024fut,TASEH:2022vvu,APEX:2024jxw,SHANHE:2023kxz,Schneemann:2023bqc,Cervantes:2022gtv,Kang:2024slu,Dixit:2020ymh,Nguyen:2019xuh,Adair:2022rtw,Ahyoune:2024klt,Grenet:2021vbb} provide strong resonant enhancement, but their characteristic volume decreases rapidly as the target frequency increases~\cite{Stern:2015kzo,Bae:2024kmy}, making it progressively harder to maintain both sensitivity and practical tunability near the sub-meV range. Open radiative concepts such as dish antennas~\cite{Horns:2012jf} retain large area and become naturally attractive at higher frequency, but sacrifice the field buildup associated with resonant confinement. More elaborate designs, such as dielectric haloscopes~\cite{Jaeckel:2013eha,Caldwell:2016dcw,Baryakhtar:2018doz,MADMAX:2024jnp,Wei:2025kgb}, can interpolate between these limits through controlled interference, though at the price of many controlled interfaces and increasingly demanding tuning as the target frequency is varied. The central question is therefore whether one can construct a detector that remains geometrically open while still achieving a parametric enhancement of the electromagnetic response.

\begin{figure}[t]
    \centering
    \includegraphics[width=8.5cm]{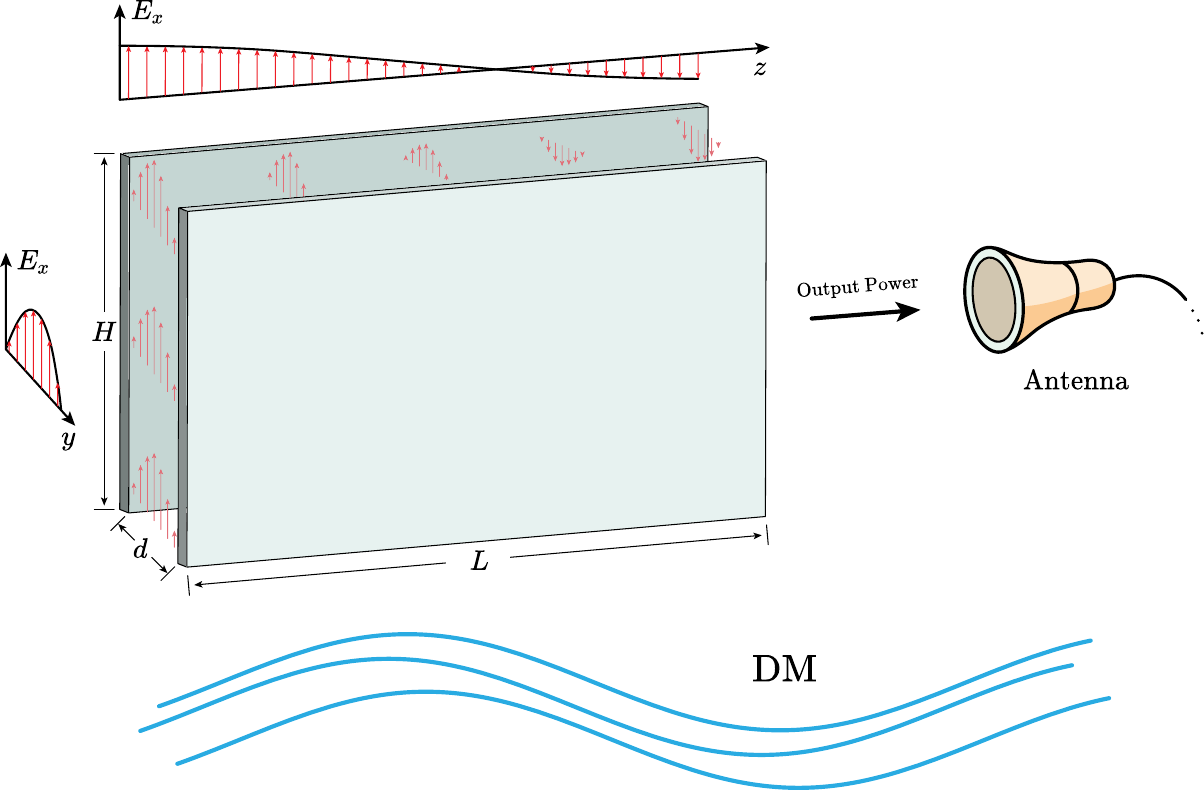}
    \caption{Near-cutoff parallel-plate waveguide haloscope. Two conducting plates separated by $d$ define a guide of width $H$ and length $L$. The electric-field profile of the lowest odd $\TE{1}$ mode is shown.}
    \label{fig:waveguide}
\end{figure}

In this Letter, we propose a haloscope based on a near-cutoff parallel-plate waveguide for $\mu{\rm eV}$-meV ultralight bosonic DM, shown schematically in Fig.~\ref{fig:waveguide}. Two conducting plates separated by a distance $d$ define a guided channel in which the lowest odd $\TE{1}$ mode can be tuned close to cutoff by adjusting the plate spacing, while the transverse width remains macroscopic. 
Rather than operating as a closed resonator, it functions as a guided-wave transducer that retains geometric openness at high frequency while avoiding the usual volume penalty of a closed cavity haloscope.
At the same time, the waveguide provides an internal field enhancement absent in a purely open radiative geometry.
Near cutoff, the guided mode acquires a small group velocity, equivalently a longer effective interaction time within the guide, while the coherent DM drive allows the guided-mode field to build up coherently along the guide. The signal therefore combines macroscopic open-area scaling with slow-wave enhancement and coherent buildup of the guided-mode field. In practice, the achievable enhancements are limited by the longitudinal coherence of the DM field, ohmic wall losses, and the detuning tolerance required to remain sufficiently close to cutoff.

We take dark photon DM as the primary benchmark, and find that a copper waveguide can reach kinetic mixing at the $10^{-15}$ level in the sub-meV window. Lower-loss implementations can further improve the sensitivity. The same architecture extends directly to axion searches in an external static magnetic field and can approach the KSVZ~\cite{Kim:1979if, Shifman:1979if} and DFSZ~\cite{Dine:1981rt, Zhitnitsky:1980tq} target region. The near-cutoff waveguide thus provides a sensitive and scalable platform for future searches for sub-meV DM.

Throughout, we work in natural units with $\hbar=c=1$.

\section{Near-Cutoff Transduction}

For dark photon DM, the background enters Maxwell's equations as an effective oscillatory current~\cite{Chaudhuri:2014dla}
\begin{equation}
\mathbf J_{\rm eff}(\mathbf r)=\varepsilon\,m_{A'}^2\,\mathbf A'(\mathbf r),
\label{eq:jeff_dpdm}
\end{equation}
with oscillation frequency $\omega\simeq m_{A'}$. We consider a parallel-plate waveguide with conducting walls at $y=0$ and $y=d$, width $H$ along $x$, and length $L$ along $z$. In the wide-guide limit, the modes are independent of $x$. The guided modes are labeled by a transverse index $j$ and polarization family, with propagation constant
\begin{equation}
\beta_j^2=\omega^2-\kappa_j^2,
\qquad
\kappa_j=\frac{j\pi}{d},
\label{eq:dispersion}
\end{equation}
so cutoff occurs at $\omega=\kappa_j$. We focus on the lowest odd $\TE{1}$ mode, whose electric field is polarized along $\hat{\mathbf x}$ and therefore couples directly to an $x$-polarized dark photon drive.

The physical mechanism is that, near cutoff, the relevant guided mode acquires a small longitudinal wavenumber and hence a small group velocity, so a spatially distributed DM source can drive an enhanced guided response without relying on a high-$Q$ standing-wave resonance. 
Expanding the fields in guided modes and applying Lorentz reciprocity, as summarized in Appendix~\ref{app:driven_mode}, gives a coupled-mode equation. For a single approximately uniform dark photon plane-wave component $\mathbf J_{\rm eff}=\mathbf J_0 e^{ik_{A'}z}$, the forward modal amplitude $a_{\TE j}^{+}$ of the $\TE{j}$ mode obeys
\begin{equation}
\frac{d a_{\TE j}^{+}}{dz}
=
s_{\TE j}\,e^{i(k_{A'}-\beta_j)z}.
\end{equation}
Here $s_{\TE j}$ is the source--mode overlap integral. It measures how strongly the effective current drives the $\TE{j}$ mode through the matching of the current polarization and transverse profile to those of the mode. For the plane-wave source above, this overlap gives
\begin{equation}
s_{\TE j}
=
-\,\frac{2\omega}{d\,\kappa_j\,\beta_j}\,J_{0,x},
\qquad (j\ {\rm odd}),
\label{eq:source_overlap_te}
\end{equation}
while even-$j$ modes vanish by symmetry. Equation~\eqref{eq:source_overlap_te} shows directly that the driven response is enhanced as $1/\beta_j$ near cutoff. Since $v_g=d\omega/d\beta_j=\beta_j/\omega$, this is equivalently a $1/v_g$ enhancement. The near-cutoff gain is therefore a \emph{slow-wave effect}, rather than a high-$Q$ cavity resonance. For the lowest mode, this enhancement can be quantified by the dimensionless slow-wave factor,
$
\mathcal B_{\rm sw} \equiv m_{A'}/|\beta_1|.
$

This slow-wave enhancement is cut off by ohmic wall loss. With finite conductivity, the propagation constant becomes complex, $ \beta_j^{\rm loss} = \Re \beta_j^{\rm loss}+i\,\Im \beta_j^{\rm loss} $.
A boundary-condition perturbative treatment~\cite{Jackson:1998nia} of the $\TE{j}$ mode gives
\begin{equation}
\left(\beta_j^{\rm loss}\right)^2
\simeq
\beta_j^2+\frac{4(1+i)\kappa_j^2R_s}{\omega \,d \, Z_0}.
\label{eq:beta_loss}
\end{equation}
where $R_s$ is the wall surface resistance and $Z_0 = 377\,\Omega$ is the vacuum impedance. See Appendix~\ref{app:lossy_modes} for details.
The second term remains finite at cutoff, so the physical wavenumber does not vanish even when the ideal lossless value $\beta_j\to 0$. Equivalently, tuning closer to cutoff reduces the phase velocity but simultaneously increases attenuation. The formal slow-wave factor $m_{A'}/|\beta_j|$ is therefore replaced by the finite quantity $m_{A'}/|\beta_j^{\rm loss}|$, and the enhancement saturates once the loss term in Eq.~\eqref{eq:beta_loss} becomes comparable to $\beta_j^2$.
In addition, the halo momentum spread sets a typical lossless wavenumber scale $\beta_j\sim m_{A'}v$. Combining this velocity-dispersion limit with the loss cutoff from Eq.~\eqref{eq:beta_loss} gives
\begin{equation}
\mathcal B_{\rm sw}
=
\frac{m_{A'}}{|\beta_1^{\rm loss}|}
\lesssim
\min\!\left[ v^{-1},\sqrt{\frac{\pi Z_0}{4\sqrt{2}\,R_s}} \, \right],
\label{eq:mbeta_bound}
\end{equation}

A second enhancement arises because the DM-induced source is distributed over the full guide length. In the coherent limit, the guided-mode amplitude $a_{\TE j}^{+}$ builds up linearly along $L$, so the mode amplitude scales as $L$ and the emitted power scales as $L^2$.
After averaging over halo velocities, the finite-length power buildup is reduced by a longitudinal coherence factor $\mathcal C_1$. When $L$ lies within the DM longitudinal coherence length, the drive remains nearly phase matched across the guide and $\mathcal C_1\simeq1$.
It is therefore useful to define the dimensionless coherence boost of the mode amplitude as
$
\mathcal B_{\rm coh}\equiv (m_{A'}L)\,\mathcal C_1 .
$

The coherent boost is limited by the effective accumulation length of the DM drive along the guide. Longitudinal buildup requires both phase coherence with the guided mode and weak attenuation over the device length. Parametrically, these conditions are
$L\lesssim|m_{A'}v-\Re\beta_1^{\rm loss}|^{-1}$
and
$L\lesssim (\Im\beta_1^{\rm loss})^{-1}$.
The length should also remain within the coherence length of the DM field, $L\lesssim(m_{A'}v)^{-1}$. Thus,
\begin{equation}
\mathcal B_{\rm coh}
\lesssim
\min\!\left[
v^{-1},
\frac{m_{A'}}{|m_{A'}v-\Re\beta_1^{\rm loss}|},
\frac{m_{A'}}{\Im\beta_1^{\rm loss}}
\right].
\label{eq:coherence_boost}
\end{equation}
Equations~\eqref{eq:mbeta_bound} and \eqref{eq:coherence_boost} therefore identify the two guided-wave enhancements of the accumulated modal response: the slow-wave response and the coherent buildup.

\section{Output Power and Sensitivity}

To estimate the observable signal, we relate the amplitude $a_{\TE 1}^{+}$ to the outgoing guided power. The modal amplitude contains both the slow-wave response and coherent accumulation, but the power flux of a near-cutoff guided mode is reduced by its group velocity, so only one net slow-wave factor remains. Similarly, the coherent $L^2$ power scaling is organized as the geometric factor $HL$ times the residual coherent enhancement $\mathcal B_{\rm coh}$. This gives the output power,
\begin{equation}
P_{\TE{1}}
=
\frac{2}{\pi}\,
\mathcal G_1\,\xi_x\,\varepsilon^2 \rho_{\rm DM}\,(HL)\,
\mathcal B_{\rm sw}\,\mathcal B_{\rm coh}\,
\frac{\Re\beta_1^{\rm loss}}{|\beta_1^{\rm loss}|},
\label{eq:power_te1}
\end{equation}
where $\mathcal G_1$ absorbs end reflections, and $\xi_x$ is the dark photon polarization fraction along $\hat{\mathbf x}$. Equation~\eqref{eq:power_te1} makes the central scaling transparent: the device preserves the macroscopic open-area scaling $(HL)$ of an open geometry, while the guided-wave enhancement is controlled by the product $\mathcal B_{\rm sw}\mathcal B_{\rm coh}$.

To bracket realistic performance, we distinguish two material benchmarks. An optimistic superconducting benchmark represents the materials-limited upper sensitivity, and is especially natural for dark photon operation, where no strong static magnetic field is required. Superconducting implementations can also be relevant to axion haloscopes, but in that setting the achievable loss becomes more sensitive to material choice and field orientation. By contrast, our default benchmark is conservative cryogenic copper, which provides a robust baseline for generic high-frequency operation and remains directly compatible with the strong magnetic-field environment required for axion searches.

For the conservative copper benchmark adopted below, we take $R_s\sim7\,\mathrm{m}\Omega$~\cite{Cahill:2016uui} at low temperature. 
The enhancement factors can be calculated as $\mathcal B_{\rm coh} \sim 230$ and $\mathcal B_{\rm sw}\sim 173$, corresponding to a total guided-wave enhancement $\mathcal B_{\rm sw}\mathcal B_{\rm coh}\sim 4\times10^{4}$. 
The choice $\mathcal B_{\rm coh}\sim230$ corresponds to a meter-scale guide in the sub-meV regime, large enough to realize substantial coherent buildup while remaining compatible with finite-length and fabrication tolerances. In the optimistic superconducting limit, one may instead approach $\mathcal B_{\rm sw}\sim10^3$ and $\mathcal B_{\rm coh}\sim10^3$, so that the total guided-wave enhancement reaches $\mathcal B_{\rm sw}\mathcal B_{\rm coh}\sim10^6$.

For comparison, a conventional resonator packages its enhancement into the cavity quality factor, defined in terms of the stored electromagnetic energy $U$ and the time-averaged dissipated power $P_{\rm diss}$ as
\begin{equation}
Q_{\rm cav}\equiv \omega\,\frac{U}{P_{\rm diss}},
\end{equation}
so that the present waveguide haloscope may be assigned an effective cavity-like enhancement
\begin{equation}
Q_{\rm wg}\sim \mathcal B_{\rm sw}\mathcal B_{\rm coh}.
\end{equation}
The copper and superconducting benchmarks therefore correspond to $Q_{\rm wg}\sim4\times10^{4}$ and $Q_{\rm wg}\sim10^{6}$, respectively, spanning the range from moderate- to high-$Q$ microwave cavities. The physical origin, however, is different: here the gain arises from slow-wave enhancement and distributed coherent buildup, rather than from standing-wave energy storage.

To translate the predicted output power into a projected reach, we compare $P_{\TE{1}}$ with the detectable signal power set by the readout. For a signal power $P_{\rm sig}$ measured with system noise temperature $T_{\rm sys}$, integration time $t$, and analysis bandwidth $\Delta\nu$, the radiometer estimate gives
\begin{equation}
{\rm SNR}
=
\frac{P_{\rm sig}}{k_B T_{\rm sys}}
\sqrt{\frac{t}{\Delta\nu}} .
\end{equation}
For a representative mass $m_{A'}=0.1\,\mathrm{meV}$, the virial linewidth is $\Delta\nu\simeq24.2\,\mathrm{kHz}$, giving $P_{\rm sig}\simeq3.6\times10^{-22}\,\mathrm{W}$ for ${\rm SNR}=5$, $T_{\rm sys}=10\,\mathrm{K}$, and $t=1\,\mathrm{day}$.

\begin{table*}[t]
\caption{Benchmark assumptions used to anchor the projected dark photon reach. Entries in the third column are listed as Conservative / Optimal, where the optimal values correspond to the lower-loss superconducting scenario.}
\label{tab:benchmark}
\begin{ruledtabular}
\begin{tabular}{llll}
Parameter & Symbol & Benchmark value & Role in the projection \\
\hline
Reference mass scale & $m_{A'}$ & $0.1\,\mathrm{meV}$ & the benchmark operating point \\
Guide width & $H$ & $1\,\mathrm{m}$ & meter-scale open collecting area \\
Threshold power & $P_{\rm sig}$ & $3.6\times10^{-22}\,\mathrm{W}$ & illustrative readout target \\
Coherent length factor & $\mathcal B_{\rm coh}$ & $230 / 10^3$ & finite-length coherent accumulation \\
Slow-wave factor & $\mathcal B_{\rm sw}$ & $173 / 10^3$ & near-cutoff guided enhancement \\
Material Surface Resistance & $R_s$ & $\sim 7\,\mathrm{m}\Omega$ / $< 0.2\,\mathrm{m}\Omega$ & wall-loss scale \\
DPDM polarization factor & $\xi_x$ & $1/3$ & average projection onto $\hat{\mathbf x}$ \\
\end{tabular}
\end{ruledtabular}
\end{table*}

\begin{figure}[t]
    \centering
    \includegraphics[width=8.5cm]{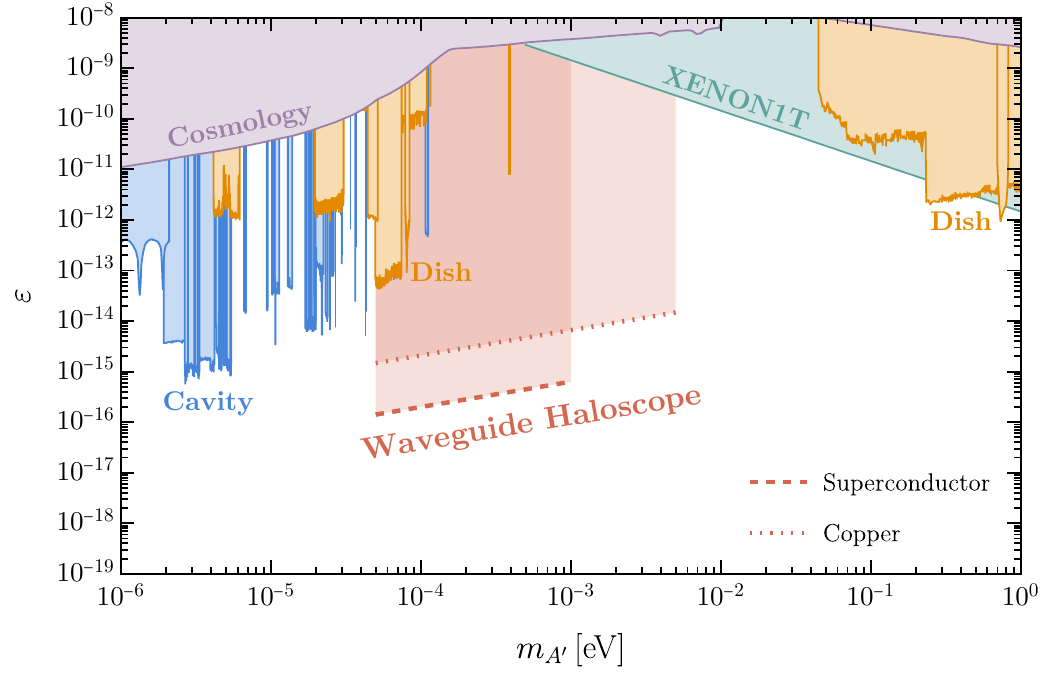}
    \caption{Projected dark photon reach of the near-cutoff waveguide haloscope. The dotted (dashed) curve is the conservative copper (optimistic superconductor) benchmark defined in Table~\ref{tab:benchmark}. Existing constraints from cavity~\cite{ADMX:2009iij, ADMX:2018gho, ADMX:2019uok, ADMX:2021nhd, ADMX:2024xbv, ADMX:2025vom, ADMX:2018ogs, Brubaker:2016ktl, HAYSTAC:2018rwy, HAYSTAC:2020kwv, HAYSTAC:2023cam, HAYSTAC:2024jch, Lee:2020cfj, Jeong:2020cwz, CAPP:2020utb, Lee:2022mnc, Yoon:2022gzp, Kim:2022hmg, Yi:2022fmn, Yang:2023yry, Kim:2023vpo, CAPP:2024dtx, Bae:2024kmy, Adair:2022rtw, Quiskamp:2022pks, Quiskamp:2023ehr, Quiskamp:2024oet, Alesini:2019ajt, Alesini:2020vny, QUAX:2023gop, QUAX:2024fut, TASEH:2022vvu, Kang:2024slu, APEX:2024jxw, Schneemann:2023bqc, Dixit:2020ymh, Cervantes:2022yzp, Cervantes:2022gtv, SHANHE:2023kxz, Nguyen:2019xuh}, dish~\cite{Suzuki:2015sza, Knirck:2018ojz, Tomita:2020usq, Brun:2019kak, DOSUE-RR:2022ise, DOSUE-RR:2023yml, An:2022hhb, Ramanathan:2022egk, Bajjali:2023uis, BREAD:2023xhc, An:2024kls, MADMAX:2024jnp, Chiles:2021gxk, Wei:2025kgb}, Xenon1T~\cite{An:2020bxd} and cosmology~\cite{Arias:2012az} are shown. The data are taken from Ref.~\cite{AxionLimits}.}
    \label{fig:dp_limit}
\end{figure}

Table~\ref{tab:benchmark} collects the benchmark assumptions used in the projected-reach figures, including both the conservative copper baseline and the optimistic lower-loss superconducting variant. The corresponding dark photon reach is shown in Fig.~\ref{fig:dp_limit}. For the copper benchmark, the projected sensitivity reaches $\varepsilon \simeq 2.1\times10^{-15}$ at $m_{A'}\simeq10^{-4}\,\mathrm{eV}$ and remains at the $10^{-15}$ level through the sub-meV range. This is the main phenomenological result: even without invoking an extreme materials scenario, the near-cutoff waveguide already probes parameter space beyond existing dish-antenna limits in the same mass window and extends laboratory sensitivity into the difficult $10^{-4}$--$10^{-3}\,\mathrm{eV}$ regime. The dotted copper curve rises gradually toward larger $m_{A'}$ because the benchmark keeps $\mathcal B_{\rm coh}\sim230$ fixed, so the physical guide length scales as $L\propto m_{A'}^{-1}$ and the geometric factor $(HL)$ in Eq.~\eqref{eq:power_te1} correspondingly decreases. The dashed curve represents the same guided behavior under a more optimistic superconducting materials scenario, in which lower wall loss extends the reach to significantly smaller kinetic mixing. Fig.~\ref{fig:dp_limit} should therefore be read as evidence that a distributed near-cutoff transducer can become competitive precisely in the high-frequency / sub-meV window while remaining macroscopically open.

\section{Axion Reach and Outlook}
The same guided transducer extends directly to axion DM in the presence of a static magnetic field $\mathbf B_0$. The axion background induces the effective current
\begin{equation}
\mathbf J_{\rm eff}^{(a)}
=
-\,i\,g_{a\gamma\gamma}\,m_a\,a_0\,\mathbf B_0,
\qquad
a_0=\frac{\sqrt{2\rho_{\rm DM}}}{m_a},
\label{eq:jeff_axion}
\end{equation}
where we used $\omega\simeq m_a$ and neglected the spatial-gradient term for nonrelativistic axion DM. The same guided-mode calculation therefore carries over after identifying the effective coupling as
\begin{equation}
\varepsilon_{\rm eff}^{(a)} \equiv \frac{g_{a\gamma\gamma}B_0}{m_a} ,
\label{eq:axion_replacement}
\end{equation}
which makes the dark photon and axion benchmarks directly comparable.
For $\mathbf B_0\parallel \hat{\mathbf x}$, the same near-cutoff $\TE{1}$ mode remains the operative readout channel.

\begin{figure}[t]
    \centering
    \includegraphics[width=8.5cm]{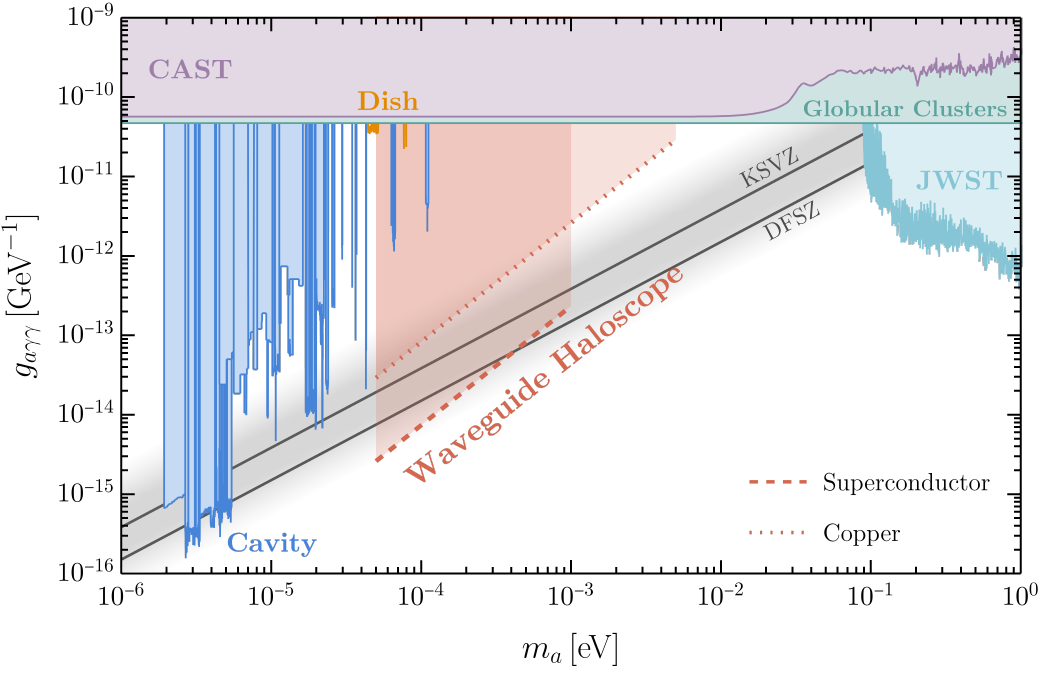}
    \caption{Projected axion reach of the near-cutoff waveguide haloscope. The dotted (dashed) curve is the conservative copper (optimistic superconductor) benchmark corresponding to the same device size used for Fig.~\ref{fig:dp_limit}. 
    Existing constraints from cavity~\cite{DePanfilis:1987dk, Hagmann:1990tj, ADMX:2009iij, ADMX:2018gho, ADMX:2019uok, ADMX:2021nhd, ADMX:2024xbv, ADMX:2025vom, ADMX:2018ogs, ADMX:2021mio, Brubaker:2016ktl, HAYSTAC:2018rwy, HAYSTAC:2020kwv, HAYSTAC:2023cam, HAYSTAC:2024jch, Lee:2020cfj, Jeong:2020cwz, CAPP:2020utb, Lee:2022mnc, Yoon:2022gzp, Kim:2022hmg, Yi:2022fmn, Yang:2023yry, Kim:2023vpo, Bae:2024kmy, CAPP:2024dtx, McAllister:2017lkb, Quiskamp:2022pks, Quiskamp:2023ehr, Quiskamp:2024oet, TASEH:2022vvu, Grenet:2021vbb, Adair:2022rtw, Alesini:2019ajt, Alesini:2020vny, QUAX:2023gop, QUAX:2024fut, CAST:2020rlf, Ahyoune:2024klt}, dish~\cite{MADMAX:2024sxs, GigaBREAD:2025lzq}, CAST~\cite{CAST:2017uph, CAST:2024eil}, James Webb Space Telescope~\cite{Pinetti:2025owq}, and globular clusters~\cite{Dolan:2022kul}, as well as the KSVZ~\cite{Kim:1979if, Shifman:1979if} and DFSZ~\cite{Dine:1981rt, Zhitnitsky:1980tq} models are also shown. The data are taken from Ref.~\cite{AxionLimits}.}
    \label{fig:axion_limit}
\end{figure}

Fig.~\ref{fig:axion_limit} shows that the axion extension is not merely a formal mapping. For the conservative copper benchmark and $B_0=8\,\mathrm{T}$, the projected reach attains $g_{a\gamma\gamma}\simeq 8.2\times10^{-14}\,\mathrm{GeV}^{-1}$ at $m_a\simeq10^{-4}\,\mathrm{eV}$, approaching the canonical QCD axion benchmark region.
The lower-loss superconducting variant extends the reach to smaller couplings, motivated by demonstrated superconducting surface resistances below $0.2\,\mathrm{m}\Omega$ in strong magnetic fields~\cite{Romanov:2020epk,Ahn:2021fgb}.
In the optimistic superconducting benchmark, the projected sensitivity reaches well into the canonical QCD axion benchmark region in the sub-meV range, covering the KSVZ benchmark over a substantial mass interval and entering the more weakly coupled DFSZ region.
The main implication is that the apparatus can serve as a viable sub-meV QCD axion haloscope platform: once a static conversion field is added, the same guided near-cutoff transducer approaches canonical QCD axion benchmark couplings under conservative assumptions, while lower-loss implementations can reach into the KSVZ--DFSZ region.

\section{Conclusion and Discussion}
We have proposed a near-cutoff parallel-plate waveguide haloscope for sub-meV bosonic DM. The key point is that the device remains geometrically open while gaining two guided-wave enhancements: the slow-wave response and the coherent buildup along the guide. For dark photons, this mechanism yields projected sensitivity below existing limits in the sub-meV range under conservative copper assumptions, with further reach possible in lower-loss implementations. With an applied static magnetic field, the same transducer also becomes an axion haloscope and can reach well-motivated QCD axion parameter space.

The remaining challenges are therefore mainly experimental rather than conceptual. 
Wall loss, cutoff detuning, end reflections, and mode matching directly determine the attainable guided-wave boost, and can be calibrated with conventional microwave techniques before a DM search is performed. A first-generation experiment can thus test the central mechanism by measuring the loss-limited slow-wave boost and coherent boost in a cryogenic copper guide, while lower-loss materials provide a clear path toward stronger dark-photon reach. For axions, the additional requirement is a high-field magnetic environment, where the same guided transducer can convert the static-field-assisted axion drive into the near-cutoff mode.

\section*{Acknowledgments}
We thank Wenyu Wang and Lei Wu for helpful discussions. 
Chuan-Yang Xing was supported by the Fundamental Research Funds for the Central Universities (No.~24CX06048A).
Bin Zhu was supported by the Shandong Province Young Scientists  
Fund Project (class A) (No.~ZR2025QA20) and the National Natural Science Foundation of China (NNSFC) under grant
No.~12275232. 

\appendix

\section{Parallel-Plate Waveguide Eigenmodes}
\label{app:eigenmodes}

We label the guided modes by $\lambda_j$, with $\lambda=\mathrm{TE},\mathrm{TM}$ denoting the polarization family and $j$ the transverse order.
The source-free forward and backward waves are written as a product of forward and backward travelling waves with transverse profiles,
\begin{subequations}
\label{eq:profiles_def_app}
\begin{align}
\mathbf E_{\lambda_j}^{\pm}(y,z)
&=
\mathbf e_{\lambda_j}^{\pm}(y)\,e^{\pm i\beta_j z},
\label{eq:profiles_def_app_E}
\\
\mathbf H_{\lambda_j}^{\pm}(y,z)
&=
\mathbf h_{\lambda_j}^{\pm}(y)\,e^{\pm i\beta_j z}.
\label{eq:profiles_def_app_H}
\end{align}
\end{subequations}
The transverse profiles satisfy the transverse Helmholtz equation,
\begin{subequations}
\label{eq:transverse_helmholtz_app}
\begin{align}
\left(\partial_y^2+\kappa_j^2\right)\mathbf e_{\lambda_j}^{\pm}(y)
&=\mathbf 0,
\label{eq:transverse_helmholtz_app_e}
\\
\left(\partial_y^2+\kappa_j^2\right)\mathbf h_{\lambda_j}^{\pm}(y)
&=\mathbf 0.
\label{eq:transverse_helmholtz_app_h}
\end{align}
\end{subequations}
where the differential operator acts on each Cartesian component.
The plate boundary conditions discretize the transverse wavenumbers,
\begin{equation}
\kappa_j=\frac{j\pi}{d},
\qquad
j=1,2,\dots,
\label{eq:kappa_parallel_plate_app}
\end{equation}
and the propagation constants satisfy the dispersion relation
\begin{equation}
\beta_j^2=\omega^2-\kappa_j^2.
\label{eq:beta_dispersion_app}
\end{equation}
Modes propagate for $\omega>\kappa_j$ and become evanescent for $\omega<\kappa_j$.
Cutoff occurs at $\omega=\kappa_j$, equivalently $\beta_j=0$.

\textit{TE modes.}
Transverse-electric modes satisfy $E_z=0$.
A convenient choice for the forward-propagating TE$_j$ transverse profiles is
\begin{subequations}
\label{eq:TEj_profiles_app}
\begin{align}
\mathbf e_{\mathrm{TE}_j}^{+}(y)
&=
\hat{\mathbf x}\,\sin\!\left(\kappa_j y\right),
\\
\mathbf h_{\mathrm{TE}_j}^{+}(y)
&=
\hat{\mathbf y}\,\frac{\beta_j}{\omega}\,\sin\!\left(\kappa_j y\right)
\;+\;
i\,\hat{\mathbf z}\,\frac{\kappa_j}{\omega}\,\cos\!\left(\kappa_j y\right).
\end{align}
\end{subequations}
The backward profiles are obtained by $\beta_j\to-\beta_j$.

\textit{TM modes.}
Transverse-magnetic modes satisfy $H_z=0$.
One convenient choice for the forward-propagating TM$_j$ transverse profiles is
\begin{subequations}
\label{eq:TMj_profiles_app}
\begin{align}
\mathbf h_{\mathrm{TM}_j}^{+}(y)
&=
\hat{\mathbf x}\,\cos\!\left(\kappa_j y\right),
\\
\mathbf e_{\mathrm{TM}_j}^{+}(y)
&=
-\,\hat{\mathbf y}\,\frac{\beta_j}{\omega}\,\cos\!\left(\kappa_j y\right)
\;+\;
i\,\hat{\mathbf z}\,\frac{\kappa_j}{\omega}\,\sin\!\left(\kappa_j y\right).
\end{align}
\end{subequations}
Again, the backward profiles are obtained by $\beta_j\to-\beta_j$.

Parallel plates also support a TEM mode with $\kappa_j=0$ and no cutoff.
We focus on the TE and TM families, since our signal enhancement arises from operation near cutoff.

\section{Driven-Mode Amplitude}
\label{app:driven_mode}

The total field driven by an effective current can be expanded in terms of the guided modes as
\begin{subequations}
\label{eq:modal_expansion_app}
\begin{align}
\mathbf E
&=
\sum_{\lambda,j}
\left[
a_{\lambda_j}^+(z)\,\mathbf e_{\lambda_j}^+(y)e^{i\beta_j z}
+
a_{\lambda_j}^-(z)\,\mathbf e_{\lambda_j}^-(y)e^{-i\beta_j z}
\right],
\\
\mathbf H
&=
\sum_{\lambda,j}
\left[
a_{\lambda_j}^+(z)\,\mathbf h_{\lambda_j}^+(y)e^{i\beta_j z}
+
a_{\lambda_j}^-(z)\,\mathbf h_{\lambda_j}^-(y)e^{-i\beta_j z}
\right],
\end{align}
\end{subequations}
where the sum runs over the TE and TM families and over transverse orders $j\ge 1$.
The symplectic orthogonality relation is
\begin{equation}
\frac{1}{2}
\int
\left(\mathbf e_{\lambda'_{j'}}^\pm\times\mathbf h_{\lambda_j}^\mp\right)\cdot\hat{\mathbf z}\,
\dd x\,\dd y
=
\pm \mathcal N_{\lambda_j}\delta_{\lambda\lambda'}\delta_{jj'},
\label{eq:orthogonality_app}
\end{equation}
where $\mathcal N_{\lambda_j}$ is a normalization factor.

To determine the modal amplitudes, apply Lorentz reciprocity to the driven fields and to a reciprocal source-free mode
$(\mathbf E_{\lambda_j}^\mp,\mathbf H_{\lambda_j}^\mp)$ in a thin slice of the guide.
In integral form, the reciprocity relation is
\begin{equation}
\int_{\partial V}
\left(
\mathbf E\times\mathbf H_{\lambda_j}^{\mp}
-
\mathbf E_{\lambda_j}^{\mp}\times\mathbf H
\right)\cdot\dd\mathbf S
=
\int_V
\mathbf E_{\lambda_j}^{\mp}\cdot\mathbf J_{\rm eff}\,\dd V .
\label{eq:reciprocity_integral_app}
\end{equation}
Taking $V$ to be a thin slice of the waveguide and using Eq.~\eqref{eq:orthogonality_app}, one obtains the coupled-mode equations
\begin{equation}
\frac{d a_{\lambda_j}^\pm}{dz}
=
\pm s_{\lambda_j}^\mp(z)e^{\mp i\beta_j z},
\label{eq:amplitude_ode_app}
\end{equation}
with transverse source overlap
\begin{equation}
s_{\lambda_j}^\pm(z)
\equiv
\frac{1}{4\mathcal N_{\lambda_j}}
\int
\mathbf J_{\rm eff}(x,y,z)\cdot\mathbf e_{\lambda_j}^\pm(y)\,\dd x\,\dd y .
\label{eq:source_overlap_general_app}
\end{equation}

For a single dark photon plane-wave component whose coherence length exceeds the guide dimensions, the effective current takes the form
\begin{equation}
\mathbf J_{\rm eff}(x,y,z)=\mathbf J_0 e^{ik_{A'}z},
\label{eq:plane_wave_current_app}
\end{equation}
so the only $z$ dependence in Eq.~\eqref{eq:source_overlap_general_app} is the phase factor $e^{ik_{A'}z}$, while the overlap integral depends only on the transverse profile
$\mathbf e_{\lambda_j}^{\pm}(y)$.
For the parallel-plate modes, this integral is independent of the propagation direction, so we can define a single
driving parameter $s_{\lambda_j}$ to parameterize the source terms,
\begin{equation}
s_{\lambda_j}^\pm(z)=s_{\lambda_j} e^{ik_{A'}z},
\quad
s_{\lambda_j}\equiv
\frac{1}{4\mathcal N_{\lambda_j}}
\mathbf J_0\cdot
\int
\mathbf e_{\lambda_j}^+(y)\,\dd x\,\dd y .
\label{eq:source_overlap_phase_app}
\end{equation}
For the TE family, $\mathbf e_{\TE j}^\pm=\hat{\mathbf x}\sin(\kappa_j y)$, so the overlap vanishes for even $j$.
For odd $j$, Eq.~\eqref{eq:source_overlap_phase_app} gives
\begin{equation}
s_{\TE j}
=
-\,\frac{2\omega}{d\,\kappa_j\,\beta_j}\,J_{0,x},
\qquad j\ {\rm odd}.
\label{eq:source_overlap_te_app}
\end{equation}
The $1/\beta_j$ factor is the slow-wave response.

Solving Eq.~\eqref{eq:amplitude_ode_app} over a guide of length $L$ requires the source term accumulated with the guided-mode phase, so we define
\begin{equation}
S_{\lambda_j}^\pm
\equiv
\int_0^L
s_{\lambda_j}^\pm(z)e^{\pm i\beta_j z}\,\dd z .
\label{eq:accumulated_drive_app}
\end{equation}
For the plane-wave source in Eq.~\eqref{eq:plane_wave_current_app},
\begin{equation}
S_{\lambda_j}^\pm
=
s_{\lambda_j} L\,
e^{i(k_{A'}\pm\beta_j)L/2}
\sinc\!\left[\frac{(k_{A'}\pm\beta_j)L}{2}\right].
\label{eq:accumulated_drive_exact_app}
\end{equation}
The coherent limit used in the main text is
$|k_{A'}\pm\beta_j|L\ll1$, for which
\begin{equation}
S_{\lambda_j}^+\simeq S_{\lambda_j}^-\simeq S_{\lambda_j}
\equiv s_{\lambda_j}L .
\label{eq:coherent_drive_app}
\end{equation}
This is the origin of the amplitude-level coherent buildup proportional to $L$.

The endpoint amplitudes are fixed by the reflection conditions
\begin{equation}
a_{\lambda_j}^+(0)=\Gamma_{\lambda_j}^0 a_{\lambda_j}^-(0),
\qquad
a_{\lambda_j}^-(L)=\Gamma_{\lambda_j}^L a_{\lambda_j}^+(L).
\label{eq:reflection_bc_app}
\end{equation}
Integrating Eq.~\eqref{eq:amplitude_ode_app} from $0$ to $L$ and imposing Eq.~\eqref{eq:reflection_bc_app} gives
\begin{equation}
a_{\lambda_j}^+(L)
=
\frac{S_{\lambda_j}^-+\Gamma_{\lambda_j}^0 S_{\lambda_j}^+}{1-\Gamma_{\lambda_j}^0\Gamma_{\lambda_j}^L}.
\label{eq:endpoint_amplitude_general_app}
\end{equation}
When the source overlap is the same for the two propagation directions and the accumulated phases are coherent,
$S_{\lambda_j}^+\simeq S_{\lambda_j}^-\simeq S_{\lambda_j}$, this reduces to
\begin{equation}
a_{\lambda_j}^+(L)
\simeq
\frac{1+\Gamma_{\lambda_j}^0}{1-\Gamma_{\lambda_j}^0\Gamma_{\lambda_j}^L}\,S_{\lambda_j}.
\label{eq:endpoint_amplitude_coherent_app}
\end{equation}
For matched terminations, $\Gamma_{\lambda_j}^0,\Gamma_{\lambda_j}^L\simeq0$, the output is simply the travelling-wave amplitude generated over the guide.

For the odd TE modes, using $J_{0,x}=\varepsilon m_{A'}^2 A'_x$ and
$E_{\rm ref}\equiv i\varepsilon\omega A'_x$ 
as the reference electric field induced by dark photon DM in free space gives
\begin{equation}
S_{\TE j}
\simeq
\frac{2i}{d\,\kappa_j}
\left(\frac{\omega}{\beta_j}\right)
(m_{A'}L)\,E_{\rm ref},
\qquad j\ {\rm odd},
\label{eq:te_accumulated_drive_app}
\end{equation}
where we used the near-cutoff relation $\omega\simeq m_{A'}$.
Using this in Eq.~\eqref{eq:endpoint_amplitude_coherent_app} gives the endpoint amplitude in terms of the reference free-space field.
The two enhancement factors are therefore the slow-wave factor
$m_{A'}/|\beta_j|$ and the coherent factor $m_{A'}L$, up to the order-unity factor $2/(d\kappa_j)$ and end-reflection factor in Eq.~\eqref{eq:endpoint_amplitude_coherent_app}.

\section{Finite-Conductivity Corrections}
\label{app:lossy_modes}

Finite wall conductivity shifts the guided-mode propagation constant and sets constraints on the achievable enhancements.
We treat wall losses as a small perturbation of the boundary conditions and parameterize the conductor by its surface
impedance.

For a good conductor, the tangential fields at the metal interface satisfy the Leontovich boundary condition,
\begin{equation}
\mathbf E_t
=
Z_s\left(\hat{\mathbf n}\times\mathbf H_t\right),
\qquad
Z_s\simeq (1-i)\,\frac{R_s}{Z_0},
\label{eq:Leontovich_BC_app}
\end{equation}
where $\hat{\mathbf n}$ is the unit normal pointing from the metal into the waveguide region, $R_s$ is the surface resistance, and $Z_0$ is the vacuum impedance.

For a TE$_j$ mode, we take $E_z=0$ and write the forward longitudinal magnetic field as
$H_z^{l}(y,z)=h_z^{l}(y)\,e^{i\beta^{l} z}$.
Throughout this derivation, a superscript $l$ denotes the lossy solution with finite conductivity.
Quantities without a superscript correspond to the perfect electric conductor (PEC) limit.
In the waveguide region, $h_z^{l}(y)$ obeys the transverse eigen-equation
\begin{equation}
\left(\partial_y^2+(\kappa_j^{l})^2\right)h_z^{l}(y)=0.
\label{eq:hz_helmholtz_TE_app}
\end{equation}
To solve the eigen-equation, we rewrite the boundary condition in Eq.~\eqref{eq:Leontovich_BC_app}
in terms of the tangential fields $E_x^{l}$ and $H_z^{l}$.
Using Maxwell's equations in the waveguide one may express
\begin{equation}
E_x^{l}
=
i\,\frac{\omega}{(\kappa_j^{l})^2}\,\partial_y H_z^{l}.
\label{eq:Ex_from_Hz_app}
\end{equation}
On the two plates, the inward normal derivative satisfies
$\partial_n\equiv \hat{\mathbf n}\cdot\nabla=\pm\partial_y$ and the $x$ component of $\hat{\mathbf n}\times\mathbf H$ picks up the
same sign.
Thus, using Eq.~\eqref{eq:Ex_from_Hz_app} in Eq.~\eqref{eq:Leontovich_BC_app}, the boundary condition can be written uniformly on both surfaces,
\begin{equation}
\left.\partial_n h_z^{l}\right|_{S}
=
-\,i\,\frac{(\kappa_j^{l})^2 Z_s}{\omega}\,\left.h_z^{l}\right|_{S}.
\label{eq:TE_impedance_BC_app}
\end{equation}
Treating this as a perturbation of the PEC problem,
to leading order in $Z_s$,
we can evaluate the prefactor on the right-hand side of Eq.~\eqref{eq:TE_impedance_BC_app} at the unperturbed eigenvalue
$\kappa_j^{l} = \kappa_j$ and replace $h_z^{l} \to h_z$.
The boundary condition for the lossy field is therefore,
\begin{equation}
\left.\partial_n h_z^{l}\right|_{S}
\simeq
g_j\,\left.h_z\right|_{S},
\qquad
g_j \equiv -(1+i)\,\frac{\kappa_j^2}{\omega}\,\frac{R_s}{Z_0}.
\label{eq:PBC_TE_app}
\end{equation}

To obtain the eigenvalue shift, we start from the two equations
\begin{subequations}
\label{eq:hz_helmholtz_TE_perturbed}
\begin{align}
\left(\partial_y^2+\kappa_j^2\right)h_z(y)&=0, \\
\left(\partial_y^2+(\kappa_j^l)^2\right)h_z^l(y)&=0.
\end{align}
\end{subequations}
Multiply the complex conjugate of the first equation by $h_z^{l}$ and the second by $h_z^*$, subtract, and integrate over $y\in[0,d]$.
Using the PEC boundary condition $\partial_y h_z|_{0,d}=0$ and Eq.~\eqref{eq:PBC_TE_app} gives
\begin{equation}
\left[\kappa_j^2-(\kappa_j^{l})^2\right]\int_0^d \left|h_z\right|^2\,\dd y
=
-g_j\left[ |h_z(0)|^2+|h_z(d)|^2\right].
\label{eq:kappa_shift_general_TE_app}
\end{equation}
For parallel plates, $h_z(y)\propto \cos(\kappa_j y)$, so the ratio of the sum of boundary terms to the integral in Eq.~\eqref{eq:kappa_shift_general_TE_app} is $4/d$.
Using $(\beta_j^{l})^2=\omega^2-(\kappa_j^{l})^2$ and $\beta_j^2=\omega^2-\kappa_j^2$, Eq.~\eqref{eq:kappa_shift_general_TE_app} gives
\begin{equation}
(\beta_j^{l})^2
=
\beta_j^2
+ (1+i)\,\frac{4 \kappa_j^2 R_s}{\omega d Z_0}.
\label{eq:beta2_shift_TE_app}
\end{equation}

We define the TE attenuation constant as
\begin{equation}
\alpha_{\mathrm{TE}_j}
\equiv
\frac{2 \kappa_j^2 R_s}{\omega\,\beta_j\,d\,Z_0},
\label{eq:alpha_TE_def_app}
\end{equation}
and the corrected propagation constant is
\begin{equation}
    (\beta_j^{l})^2
    =
    \beta_j^2+2(1+i)\,\beta_j\,\alpha_{\mathrm{TE}_j}.
    \label{eq:beta_TE_correction}
\end{equation}
In the small-loss regime $\alpha_{\mathrm{TE}_j}\ll \beta_j$, one may take the square root to obtain
\begin{equation}
    \beta_j^{l} \simeq \beta_j + (1+i)\alpha_{\mathrm{TE}_j}.
\end{equation}
This implies that the amplitude of a forward wave
$e^{i\beta_j^{l} z}$ decays as $e^{-\alpha_{\mathrm{TE}_j} z}$, which motivates the definition of $\alpha_{\mathrm{TE}_j}$ as the attenuation constant.

An analogous treatment for TM$_j$ yields the same form of correction to the propagation constant as in Eq.~\eqref{eq:beta_TE_correction}.
The corresponding attenuation constant is
\begin{equation}
\alpha_{\mathrm{TM}_j}\equiv \frac{2\omega R_s}{\beta_j\,d\,Z_0}.
\label{eq:beta2_shift_TM_app}
\end{equation}

\bibliographystyle{utphys}
\bibliography{ref}

\end{document}